\documentclass[epj]{svjour}

\newif\ifpdf            
\ifx\pdfoutput\undefined \pdffalse \else \pdftrue \fi
\ifpdf 
\usepackage[pdftex]{graphicx}
\pdfcompresslevel9 \DeclareGraphicsExtensions{.jpg,.pdf,.png,.mps}
\else 
\usepackage[dvips]{graphicx}
\DeclareGraphicsExtensions{.eps,.ps} \fi

\usepackage{graphics}
\usepackage{amsmath}
\usepackage{bm}

\def\mat#1{\ensuremath{\underline{\underline{{\bm{#1}}}}}}
\def\vec#1{\bm{#1}}

\begin{document}

\title{Reflectivity of cholesteric liquid
crystals with spatially varying pitch}

\author{Samuel Kutter\thanks{\email{kutter@cantab.net}} \and
Mark Warner\thanks{\email{mw141@cam.ac.uk}}}
\institute{Cavendish Laboratory, University of Cambridge,
Madingley Road, Cambridge CB3 0HE, UK
}

\date{\today}

\abstract{Solids with spatially varying photonic structure offer
gaps to light of a wider range of frequencies than do simple
photonic systems.  We solve numerically the field distribution in
a cholesteric with a linearly varying inverse pitch (helical
wavevector) using equations we derive for the general case.  The
simple idea that the position where the Bragg condition is locally
satisfied is where reflection takes place is only true in part.
Here, reflection is due to a region where the waves are forced to
become evanescent, and the rate of variation of structure
determines over which distance the waves decay and therefore how
complete reflection is.  The approximate local Bragg-de Vries
schemes are shown to break down in detail at the edges of the gap,
and an analytical estimate is given for the transmission
coefficient.  }

\maketitle

\section{Introduction}

Photonic band structures can be made to vary spatially by varying
the lattice repeat distance as one penetrates the sample.  An
example where this technique has been used with great success is
that of cholesterics.  In such liquid crystals the director (the
direction of high refractive index) rotates on moving
perpendicularly to itself to give a helical structure, see the
sketch fig.~\ref{pitch}. These structures are known
\cite{devries51} to have a single photonic stop gap for normal
incidence of circularly polarized light of the same handedness as
the helical cholesteric and no gap for light of the opposite
handedness.

We study here variable pitch cholesterics (VPCs). The motivation
for spatially varying structures is to have reflection over a
broader range of frequencies than simply that provided by the gap
of a uniform system.  This has been usefully employed in wide band
reflectors and circular polarizers for display purposes. The idea
employed to interpret such effects is that light of a given
frequency penetrates until the repeat distance of the solid is
locally that of the wavelength of this light in the solid.  At
this point a Bragg-de Vries condition is satisfied and reflection
occurs.  Our purpose is to explore whether such an approach has
any utility and what limitations arise, for instance unwittingly
allowing unwanted transmission under some circumstances.  The
counter argument to this simple argument would be that Bragg
effects are coherent -- one needs very many such repeats to get
sharp and total Bragg interference and in a varying medium this is
achieved nowhere.  To some extent the answer must be that
significant evanescence of the wave must occur between first
encountering the gap and leaving it. We explore this numerically
and by approximate methods.  A rather different theoretical
approach has used the Berreman matrix formulation \cite{lu-li97}.

Since cholesterics are locally nematic and therefore also fluid, a
VPC must be solidified to stabilize the pitch gradient once it has
been set up. One method is to photo-induce a concentration
gradient in a blend of chiral and achiral photo-crosslinkable
mesogens \cite{Broer,Broer_99}.  A network (elastomer) is formed
which freezes in the structure. Otherwise, thermal diffusion
between two cholesteric layers with different pitches can be used
to set up a variation in pitch which can then be frozen in by
quenching to the glassy state \cite{mitov-boudet98}.

\section{Governing equations}

\begin{figure}
\centering \resizebox{0.45\textwidth}{!}
{\includegraphics{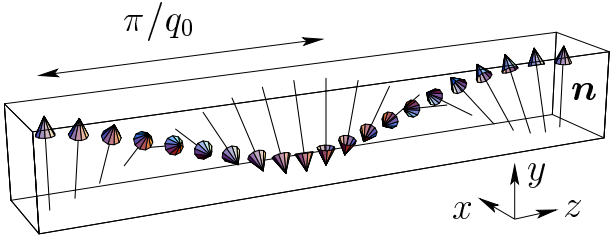}} \caption{A classical, constant pitch
cholesteric structure. The arrows indicate the nematic director
prevalent in the $xy$-plane at a certain $z$ coordinate. Note that
$\vec{n}$ and $-\vec{n}$ describe the same physical state, namely an
orientational order along the {\em direction} of $\vec{n}$ (or
$-\vec{n}$).}
\label{pitch}
\end{figure}

In conventional cholesteric liquid crystals, the nematic director
$\vec{n}$ rotates uniformly in a plane as one advances in a line
perpendicular to this plane. We shall adopt the following
convention: the nematic director $\vec{n}$ is confined to the
$xy$-plane, and the direction in which the director changes is
therefore the $z$-axis, see fig.~\ref{pitch}.  The director of
ordinary cholesteric liquid crystals subsequently becomes
\begin{equation}
    \vec{n}=\left( \cos \phi(z),\sin \phi(z),0\right)
    \quad   \mbox{with} \quad
    \phi(z)=\frac{2\pi}{p_0} z \label{cholesteric-director},
\end{equation}
where $p_0$ is the pitch length.

To allow for the most general phase evolution, we look at the
governing equation for electromagnetic waves in a cholesteric
liquid crystal whose pitch does not need to be a constant. In
other words, we do not assume the specific form $\phi=(2\pi/p_0)z$
in eq. (\ref{cholesteric-director}), but leave the phase in the
most general form $\phi(z)$. On the other hand, we restrict
ourselves to electromagnetic waves which propagate along the
$z$-axis only, i.e. perpendicular to the director $\vec{n}$.
Therefore the electric and magnetic fields of the wave only have
$x$ and $y$ components \cite{jackson75}.

The anisotropy of the liquid crystalline phase described by the director
$\vec{n}$ implies that locally the dielectric tensor respects the
uniaxial symmetry of the liquid crystalline phase. It therefore
assumes the following form:
\begin{equation}
    \mat{\epsilon}=
    \epsilon_{1}\mat{\delta}+
        \left(\epsilon_{2}-\epsilon_{1}\right)
            \vec{n}\vec{n}^{\sf T},
            \label{die-tensor}
\end{equation}
where $\mat{\delta}$ is the $3\times 3$ unit matrix and where
$\epsilon_{2}$ and $\epsilon_{1}$ are the dielectric constants
parallel and perpendicular to the director $\vec{n}$
respectively. Since $\vec{n}$ rotates in the $xy$-plane along the
$z$-axis eq.~(\ref{cholesteric-director}), the dielectric tensor
$\mat{\epsilon}$ eq. (\ref{die-tensor}) depends too on the $z$-coordinate.

In the present case, the electromagnetic wave equation (assuming a
unit magnetic permeability $\mu=1$), reads \cite{jackson75}:
\begin{equation}
    \frac{1}{c^2}\frac{\partial^2 \vec{D_t}}{\partial t^2}
    =\frac{\partial^2 \vec{E_t}}{\partial z^2},\label{t-wave-eq}
\end{equation}
with the dielectric displacement field $\vec{D_t}=\mat{\epsilon}\vec{E_t}$.

Introducing the time dependence $\vec{E_t}(z)=e^{i\omega t} \vec{E}(z)$ (and
similarly for $\vec{D}$), all time derivatives become a mere
multiplication by $i\omega$, and we obtain:
\begin{equation}
    -k_0^2 \vec{D}=\frac{\partial^2\vec{E}}{\partial z^2}
    \ \mbox{with}\
    k_0=\frac{\omega}{c}.\label{wave-eq}
\end{equation}

Note that the tensor $\mat{\epsilon}$ is in general not diagonal
in the fixed laboratory frame $(xyz)$ (cf. eqs.
(\ref{cholesteric-director}) and (\ref{die-tensor})), however due
to its symmetry given by eq.~(\ref{die-tensor}), the
dielectric tensor is diagonal in the frame defined by the rotating
director $\vec{n}(z)$. In this frame, which we denote by a tilde
and $(\xi,\eta)$ coordinates, the electric displacement field is
simply given by:
\begin{equation}
    \tilde{D}_{\xi}=\epsilon_1\tilde{E}_{\xi}
    \ \mbox{and}\
    \tilde{D}_{\eta}=\epsilon_2\tilde{E}_{\eta}.
\end{equation}
We take $\eta$ to be the direction along $\vec{n}$ and hence typically
$\epsilon_2 > \epsilon_1$.

The coordinate transform into the rotating frame (attached to the
director $\vec{n}$), reads as follows:
\begin{equation}
    \tilde{\vec{E}}=\mat{R}(z)\vec{E}
    \ \mbox{with}\
    \mat{R}(z)=
        \begin{pmatrix}
                \cos \phi(z) &    \sin \phi(z) \\
            -\sin \phi(z)    & \cos \phi(z) \\
            \end{pmatrix}.
\end{equation}
In this new frame, eq.~(\ref{wave-eq}) transforms into
\begin{equation}
    k_0^2\, \mat{\epsilon}\, \tilde{\vec{E}}=
        \begin{pmatrix}
               -\partial^2_z+\phi'          & 2\phi'\partial_z+\phi''   \\
           - (2\phi'\partial_z+\phi'')  & -\partial^2_z+\phi'       \\
            \end{pmatrix}\tilde{\vec{E}}\label{rotwave},
\end{equation}
where $\partial_z=\partial/\partial z$ and
$\phi'=\partial\phi/\partial z$, and $\mat{\epsilon}$ is now the
diagonal $2\times 2$ dielectric tensor
$\begin{pmatrix}
                \epsilon_1 &    0 \\
            0 & \epsilon_2 \\
\end{pmatrix}$ in the $(\xi, \eta)$ frame.  Dropping the tilde,
here and  henceforth, eq.~(\ref{rotwave}) reads:

\begin{eqnarray}
    k_0^2\,\epsilon_1\, E_{\xi}
        & = &
        -E_{\xi}''+E_{\xi}  \phi'^2 + 2 E_{\eta}'\phi'+E_{\eta}\phi''
            \nonumber\\
    k_0^2\,\epsilon_2\, E_{\eta}
        & = &
        -E_{\eta}''+E_{\eta}\phi'^2 - 2 E_{\xi}' \phi'-E_{\xi}\phi''.
                \label{eq-system}
\end{eqnarray}

\section{Cholesterics with constant pitch}

Ordinary cholesteric liquid crystals are characterised by a
constant pitch $p_0$. Hence the director $\vec{n}$ performs a
steady rotation as one travels in $z$-direction. The physical
system is periodic in $\frac{p_0}{2}$, although
eq.~(\ref{cholesteric-director}) is only periodic in $p_0$,
reflecting the fact that both $\vec{n}$ and $-\vec{n}$ describe
the axis of anisotropy of a uniaxial system equally well, see
eq.~(\ref{die-tensor}).

For constant pitch, the phase $\phi(z)$ is given by $\phi=q_0 z$
with $q_0=\frac{2\pi}{p_0}={\rm const.}$ and hence $\phi'=q_0$ and
$\phi''=0$ -- the classical problem of de Vries. The wave equation
(\ref{eq-system}) is exactly solvable \cite{devries51}  in the
$(\xi,\eta)$ frame using the solution:
\begin{equation}
    \vec{E}=
        \begin{pmatrix}
            A   \\
            iB   \\
            \end{pmatrix} e^{ikz}\; . \label{plane-wave}
\end{equation}
We obtain two linear homogeneous equations for the coefficients
$A$ and $B$. There can only be a non-trivial solution if the
determinant of the corresponding matrix equals zero:
\begin{equation}
    k^4
    -k^2 (k_0^2[\epsilon_1+\epsilon_2]+2 q_0^2)
    +(\epsilon_1k_0^2-q_0^2)(\epsilon_2k_0^2-q_0^2)=0
\; .    \label{det-eq}
\end{equation}

\begin{figure*}
\centering \resizebox{0.8\textwidth}{!}
{\includegraphics{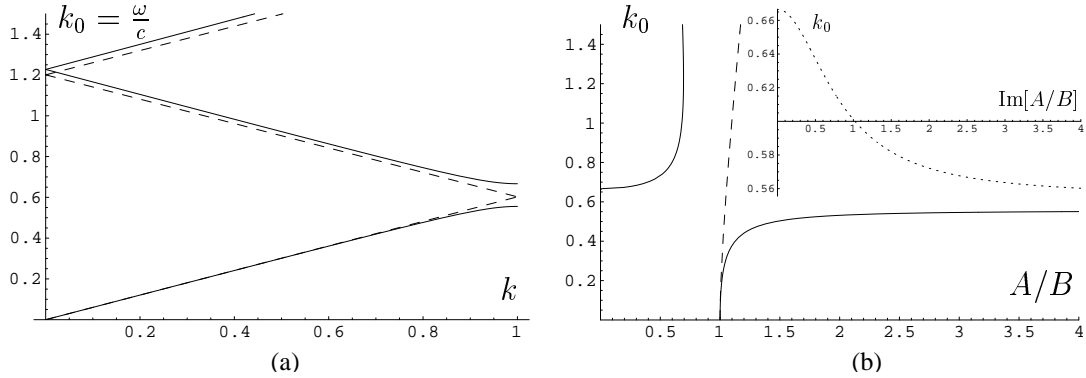}} \caption{(a) Reduced temporal
frequency $k_0=\omega/c$ vs. wave vector $k$: in a cholesteric
liquid crystal with constant pitch $p_0$, there are two pairs of
solutions.  One of them exhibits a single stop gap for normal
incidence at frequencies $k_0=\omega/c$ between
$[q_0/\sqrt{\epsilon_2},q_0/\sqrt{\epsilon_1}]$ (solid line) with
 $q_0$ being the cholesteric's wave vector $2\pi/p_0$. The other is gapless (dashed line).  The plot can be extended to
negative values of $k$ by reflection in the $k_0$-axis. In this
example, we arbitrarily chose $\epsilon_1=2.25$, $\epsilon_2=3.24$
(the dielectric constants perpendicular and parallel to the local
director) and $q_0=1$. (b) The corresponding ratio of the fields
along and perpendicular to the rotating director as a function of
frequency $k_0=\omega/c$. The dotted line of the inset shows how
the ratio $A/B$ becomes imaginary in the gap and that the
variation of $A/B$ is strong near the edges of the gap.}
\label{devries-band}
\end{figure*}

This quadratic equation in $k^2$ establishes a relationship
between the temporal frequency $\omega=k_0 c$ and the
corresponding wave vector $k$. In the reduced zone scheme, the
relationship $k_0$ as a function of $k$ is shown in
fig.~\ref{devries-band}(a).

The ratio $A/B = E_{\xi}/E_{\eta}$ of the field components also
varies as $k(k_0)$ is varied:
\begin{equation}
    \frac{A}{B}(k_0)=\frac{2kq_0}{k^2+q_0^2-\epsilon_{2}k_0^2}  \; ,
    \label{field-ratio}
\end{equation}
where $k$ and $k_0$ are related by eq.~(\ref{det-eq}), see
fig.~\ref{devries-band}(b).  The choice of the dielectric
constants $\epsilon_1$ and $\epsilon_2$ in this illustration
corresponds to refractive indices perpendicular and parallel to
$\vec{n}$ of $n_1=1.5$ and $n_2 = 1.8$.

Equation (\ref{det-eq}) allows two pairs of solutions: for each
fixed frequency $k_0=\frac{\omega}{c}$, there are two solutions
whose wave vector $k$ have the same magnitude, but  opposite sign,
reflecting the fact that there are waves travelling in the two
opposite directions along the $z$-axis. Apart from this trivial
pairing, there are two genuinely different solutions to
eq.~(\ref{det-eq}). One of them exhibits a band gap in the
frequency range $\omega/c\in
[q_0/\sqrt{\epsilon_2},q_0/\sqrt{\epsilon_1}]$. Note that due to
the periodicity in the cholesteric structure, the graphs in
fig.~\ref{devries-band}(a) were shifted by 1 along the $k$-axis to
yield the reduced zone representation.

The non-trivial pair of solutions, the de Vries eigenstates, are
in general of elliptical polarization, the principal directions of
which rotate with $\vec{n}$. At the two band edges at
$k_0=q_0/\sqrt{\epsilon_2}$ and $q_0/\sqrt{\epsilon_1}$, the
states are plane polarized with the plane of polarization rotating
with $\vec{n}$.

Light with the same handedness as the cholesteric has states
$(E_{\xi},0)$ and $(0,E_{\eta})$ at the band edges, corresponding
to $\vec{E}$ respectively along $\vec{n}$ or perpendicular to
$\vec{n}$. The ratio $A/B$ is then $\infty$ or 0 respectively. The
former is along a polarizable direction and hence produces a lower
energy that the latter even though the spatial variation $k$ is
the same --- a gap has been produced. Light rotating with the
opposite handedness does not couple coherently to the cholesteric
and retains a  gapless dispersion relation, the dashed line in
fig.~\ref{devries-band}. The de Vries problem, in this photonic
band structural language of fig.~\ref{devries-band}, can be
extended to non-uniform but periodical director rotation
\cite{bermel-warner02}.

\section{Cholesterics with varying pitch}

In varying pitch cholesterics (VPCs) the nematic director
$\vec{n}$ rotates non-uniformly along the $z$-axis, i.e., the
phase $\phi(z)$ of eq.~(\ref{cholesteric-director}) does not have
the simple form $\phi(z)=q_0 z=2\pi z/p_0$ with $q_0={\rm const}$,
but assumes a more general form. It has been reported that the
local pitch can vary exponentially with $z$ \cite{lu-li97}, i.e.,
we can write $q=q_0\exp(-\kappa z)$. Other  careful, scanning
electron microscopy investigations
 \cite{Broer,Broer_99,mitov-boudet98},
showed the pitch to be linear over a range of $z$-values. Here, we
assume the simplest possible inverse pitch variation, namely a
linear dependency on $z$, leading to a quadratic dependence of the
phase $z$:
\begin{equation}
    q=q_0+\gamma q_0^2 z
    \
    \Longrightarrow
    \
    \phi(z)=\left(q_0+\textstyle{\frac{1}{2}}\gamma q_0^2 z\right)z
           \; . \label{linear-ramp}
\end{equation}

\subsection{Local de Vries approximation}

A simple framework for considering a wide band circular polarizer,
filter or reflector now presents itself
\cite{Broer,Broer_99,mitov-boudet98}. As light of a given
frequency enters and accordingly has a given wavelength in the
medium, then since $q(z)$ changes, there will be a depth at which
a Bragg-like condition will be satisfied, and at this depth the
wave will be totally reflected. This is where the wavelength in
the medium matches the local periodicity $2\pi/q(z)$.  We  examine
how much credence to put in this description. For Bragg reflection
to occur, one requires a semi infinite medium, at least a
thickness of sample corresponding to very many wavelengths. In the
variable pitch cholesteric, the Bragg condition is met strictly
only at a point and a Bragg mechanism fails. Indeed, instead of a
total reflection, there may be circumstances where it is only
partial, as we find below.

One can take an analogous approach of applying de Vries theory
locally to argue for reflection of light of a given wavelength at
the particular point in the VPC where for light at this frequency
a de Vries gap is encountered.

In the constant pitch case, one can determine an eigen wave vector
$k$ given a frequency $\omega$ (or equivalently $k_0$), see
fig.~\ref{devries-band}. The pitch wave vector $q_0$ played the
role of a parameter in the function $k=k(\omega)$. Let us now
change perspective and consider the frequency $\omega$ or
$k_0=\frac{\omega}{c}$ as a parameter, but allow the inverse pitch
length $q$ to change according to eq.~(\ref{linear-ramp}). In the
following, we choose $k_0=\frac{\omega}{c}=2$, $\epsilon_1=2.25$,
$\epsilon_2=3.24$, $q_0=1$, and $\gamma=0.05$.

\begin{figure}
\centering \resizebox{0.4\textwidth}{!}
{\includegraphics{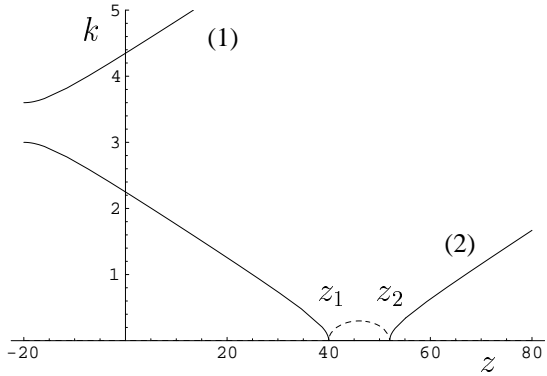}} \caption{Two of the four branches
of the local wave vector $k$ vs. $z$-coordinate: assuming a fixed
frequency $k_0$, eq.~(\ref{det-eq}) links the wave vector $k$ and
the inverse pitch length $q_0$. Previously, $q_0$ was taken to be
a constant, but eq.~(\ref{linear-ramp}) introduces a dependence on
the $z$-coordinate. Thereby, the relation between the wave vector
$k$ and the local inverse pitch length $q$ becomes a function of
$z$. For this plot, we have chosen $\epsilon_1=2.25$,
$\epsilon_2=3.24$, $k_0=2$, $q_0=1$, and $\gamma=0.05$ ($q_0^2
\gamma$ being the spatial rate of change of cholesteric
wavevector) . With this choice of parameters, the frequency $k_0$
lies below the forbidden frequency range at $z=0$. The forbidden
zone of branch (2) extends over the region
$z\in[z_1,z_2]=[40,52]$.} \label{local-wave-vec}
\end{figure}

By this reasoning, we are able to determine a wave vector $k$
satisfying the de Vries condition eq.~(\ref{det-eq}) given a
particular cholesteric inverse pitch length $q$, which in turn
depends on the $z$-coordinate by means of eq.~(\ref{linear-ramp}).
Hence, we can plot an hypothetical wave vector $k$ against the
local position $z$, fig.~\ref{local-wave-vec}. Note that for a
given frequency $k_0$, there are again four solutions derived from
eq.~(\ref{det-eq}). Figure \ref{local-wave-vec} only shows two
branches, one of them becoming imaginary between $z_1$ and $z_2$,
with
$z_{i}=\frac{1}{q_0^2\gamma}\left(\sqrt{\epsilon_{i}}k_0-q_0\right)$.
 The second pair of branches can be constructed trivially by
introducing minus signs: $k\mapsto -k$.  In the region $[z_1,z_2]$
where the frequency $k_0$ falls into the band gap for two of the
four branches, the corresponding wave vector $k$ becomes
imaginary, indicating an evanescent behaviour of the wave. We
expect there a decaying and increasing solution (see
eq.~(\ref{plane-wave}) for imaginary $k$).

One can now return to the question of reflection, or equivalently
transmission. The band gap starts at $z_1$ and extends to $z_2$.
Transmission depends on how close $z_2$ is to $z_1$ and how rapid
the exponential decay in this gap is.

\subsection{Numerical investigation}
\begin{figure}
\centering \resizebox{0.45\textwidth}{!}
{\includegraphics{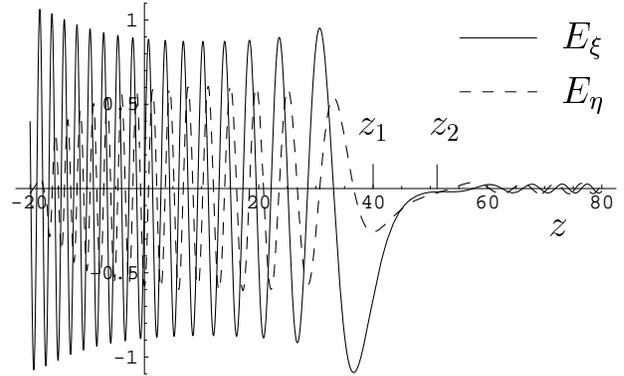}} \caption{Numerical solution of
eq.~(\ref{eq-system}) with $\phi(z)=(q_0+\frac{1}{2}\gamma q_0^2
z)z$: the initial conditions at $z=0$ are chosen such that the
solution decays over the evanescent range $z\in[z_1,z_2]=[40,52]$.
The numerical parameters are the same as in
fig.~\ref{local-wave-vec}, and the initial conditions as follows:
$E_{\xi}(0)=-0.036$, $E_{\eta}(0)=-0.58$, $E_{\xi}'(0)=2.0$, and
$E_{\eta}'(0)=-0.067$ (the fields and their gradients in the
rotating frame $(\xi,\eta)$ at $z=0$).} \label{numerical}
\end{figure}

The above tentative interpretation of the solution being locally
de Vries-like is qualitatively confirmed by looking at a full
numerical solution in fig.~\ref{numerical}. Here, we have chosen
the initial conditions at $z=0$ such that we obtain a decaying
solution in the forbidden zone as opposed to an increasing
solution, which one obtains for different initial conditions.
From the simultaneous field equations~(\ref{eq-system}) for
$E_{\xi}$ and $E_{\eta}$, it is clear that one could divide
through by a constant, say $E_{\eta}(z=0)$.  The required 4
boundary conditions then reduce to 3 choices, the 4th being thus
set to $E_{\eta} = 1$ at $z=0$, reflecting the arbitrariness of
the overall scale in the problem.

 Several features
of this solution are important for what follows. The amplitudes on
either side of the gap region differ considerably. The small
amplitude to the right of the gap after partial transmission gives
a small Poynting flux, $\vec{E}\wedge \vec{H}$, of energy in the
positive $z$ direction. Although the amplitudes on the left are
much larger, they correspond to the sum of an incident and
reflected wave resulting in a correspondingly small Poynting
vector.  The ratio $A/B$ of the field amplitudes is approximately
constant away from the band edges at $z_1$ and $z_2$.  In the de
Vries, constant pitch solution this ratio is respectively $\infty$
and $0$ at these points, extreme values that are not achieved in
this case of partial barrier penetration.  The ratio also becomes
very large at $z = - 20$ which for our value $0.05$ of $\gamma$ is
the point where there is no rotation of the director at all, that
is where $q(z) \rightarrow 0$. Locally the medium is achiral and
there is  no need for the two-component, de Vries-like solution.

One can now qualitatively analyze reflection and transmission.
With rising $q$ vector on penetrating the VPC, one must eventually
reach what would be the lower band edge in a uniform de Vries
system. However, the gap translates into a spatial region of
decaying gap states, and ultimate transmission is determined by
how wide this region is. In particular does significant decay of
the wave function occur before the propagating region of $z>z_2$
is attained? If not, then transmission is higher and reflection
lower. Decay depends therefore in part on the gap's spatial extent
\begin{equation}
z_2-z_1=\frac{k_0}{q_0^2\gamma}(\sqrt{\epsilon_2}-\sqrt{\epsilon_1})\label{space-gap}
\end{equation}
 over which decay occurs. Clearly, if the spatial
variation is weaker, i.e. $\gamma$ small, then the gap is
spatially more extended and decay has a greater length over which
to take place (see also the discussion of the transmission
coefficient leading up to eq.~(\ref{trans-coeff})).

\subsection{Approximate solutions}

For constant (inverse) pitch, the solution eq.~(\ref{plane-wave})
was a plane wave. In the case of a cholesteric with a
$z$-dependent (inverse) pitch, we choose a more general trial
function, namely a WKB-like (Wentzel, Kramers and Brillouin)
ansatz, assuming for simplicity that the two field components
$E_{\xi}$ and $E_{\eta}$ share the {\em same} phase $S(z)$:
\begin{equation}
    \begin{pmatrix}
        E_{\xi}    \\
        E_{\eta}   \\
        \end{pmatrix}
    =
    \begin{pmatrix}
        A    \\
        iB   \\
        \end{pmatrix} e^{iS(z)}. \label{wkb}
\end{equation}
Inserting eq. (\ref{wkb}) into the wave equation (\ref{eq-system})
yields two linear equations for $A$ and $B$, which have a
non-trivial solution only if the corresponding determinant is
zero. Therefore, the analogous equation to eq. (\ref{det-eq}) is:
\begin{eqnarray}
+ S''\gamma^2 q_0^2\left(i(\epsilon_1+\epsilon_2) k_0^2
    -2 i q_0^2 w^2 -q_0^2 S''\right)\nonumber&&\\
+ S'^4 q_0^4 \gamma^4 + S' 4 i q_0^4 w \gamma^2 \label{determinental}&&\\
- S'^2 \gamma^2 q_0^2 \left((\epsilon_1+\epsilon_2)k_0^2
    +2q_0^2 w^2 + 2iq_0^2 \gamma^2 S''\right)\nonumber &&\\
\epsilon_1\epsilon_2k_0^4
    -(\epsilon_1+\epsilon_2)k_0^2 q_0^2 w^2
        +q_0^4 w^4 + q_0^4\gamma^2 &=&0\nonumber
\end{eqnarray}
where we have used a rescaled variable $w=1+q_0\gamma z$. Also note
that the prime denotes the differentiation with respect to the new
variable $w$: $S'\equiv\partial S(w)/\partial w$.

This equation is not just an algebraic equation like
eq.~(\ref{det-eq}), but rather a non-linear second order
differential equation for $S$.  An approximation is buried in this
procedure. The components $A$ and $B$ were taken as constants, any
variation with $z$ they might have in common being taken up as
logarithmic terms in the general phase $S(z)$ (including any usual
pre-exponential terms like $(S')^{-1/4}$ that one would find in
conventional WKB). However, we know from eq.~(\ref{field-ratio})
and fig.~\ref{devries-band}(b) that the ratio $A/B$ in the uniform
case depends on $k$ which is now varying with respect to $q(z)$
since the pitch is spatially varying.  Hence even in a local de
Vries picture, $A/B$ must be a function of $z$ through $q(z)$.
This variation has been ignored on inserting (\ref{wkb}) into
eq.~(\ref{eq-system}). The full numerical solution,
fig.~\ref{numerical}, indeed shows the ratio $A/B$ to vary
spatially. From fig. \ref{devries-band}(b), one can see that the
de Vries value is $\infty$ at the lower edge and $0$ at the upper.
However, the variation $A/B$ goes to zero as $\gamma\rightarrow
0$, as expected, and can be estimated for small $\gamma$ to be
proportional to $\gamma k_0$, implying a small variation of $A/B$
for small $\gamma$.

Furthermore, the form (\ref{wkb}) is expected to be valid in the
regions where the variation $A/B$ is small. By looking at
fig. \ref{devries-band} with the axes interchanged, one recognizes
that the variation of $A/B$ is strong near the gap edges, but weak
otherwise.

Ignoring this potential shortcoming in the ansatz~(\ref{wkb}), and
neglecting terms involving $S''(w)$, the determinental consistency
equation~(\ref{determinental}) for $A$ and $B$ reduces to a 4th
order algebraic equation for $S'(w)$. This equation can be solved
analytically, giving a very long and complicated expression $S_1'$
for this approximation to $S'(w)$.

\begin{figure}
\centering \resizebox{0.4\textwidth}{!}
{\includegraphics{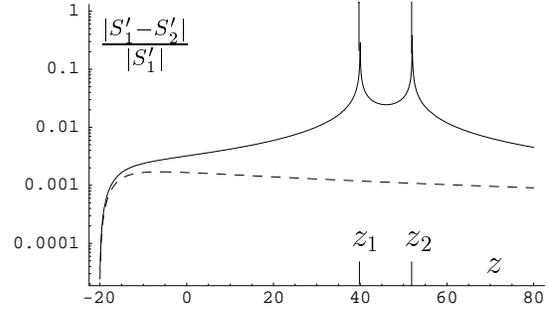}} \caption{Comparison of the two
approximations made: we plot the relative error
$(|S_1'-S_2'|)/|S_1'|$ between the two approximations $S_1'$ and
$S_2'$ over the $z$-range. The two lines correspond to the two
solutions, see the two branches in fig.~\ref{local-wave-vec}. In
the case of the branch with the forbidden zone (continuous line),
the approximation (\ref{approx}) for obtaining $S_2$ is valid a
long distance away from the points $z_1$ and $z_2$, but breaks
down in their vicinity.} \label{comparison}
\end{figure}

In fig.~\ref{comparison} we compare this solution $S_1'$ to a
further simplified approach: after switching back to the old $z$
variable, we take $\gamma\rightarrow 0$ to obtain $S_2'$; this
corresponds exactly to the de Vries wavevector function derived
from eq.~(\ref{det-eq}) and used in fig.~\ref{local-wave-vec}:
$S_2'$ stands for the local wave vector $k(z)$ of a plane wave
with frequency $\omega=k_0 c$ in a cholesteric with a inverse
pitch $q(z)$, which we consider as locally constant  -- thus
taking $\gamma = 0$ here means that we ignore derivatives of the
phase $\phi$, in other words an adiabatic approximation:
\begin{eqnarray}
    S_2'&=&\pm\Big(\textstyle{\frac{1}{2}}k_0^2 (\epsilon_1+\epsilon_2)
        +q^2 \label{approx}\\
    &&  \ \pm\textstyle{\frac{1}{2}}k_0
        \sqrt{(\epsilon_1-\epsilon_2)^2 k_0^2
            +8q^2(\epsilon_1+\epsilon_2)}\Big)^{1/2},
        \nonumber
\end{eqnarray}
where $q=q(z)$ again is given by eq.~(\ref{linear-ramp}).

We note that, for the branch with the forbidden zone, the
approximation which takes us from $S_1'$ to $S_2'$ is valid far
away from the points $z_1$ and $z_2$, where the phase variation is
zero on the transition from being real to becoming imaginary or
vice versa. However, close to these points, approximating $S_1'$
with $S_2'$ clearly breaks down, as fig.~\ref{comparison} shows.
In any event at the band edges where $S'(z) \rightarrow 0$ the
WKB-like approach breaks down for the usual reasons.

\subsection{Comparison of approximate schemes with numerical solutions}

So far, we have only compared the validity of two consecutive
approximations $S_1'$ and $S_2'$ with each other. However, the
true test is to compare the approximation to the full numerical
solution.
\begin{figure*}
\centering \resizebox{0.85\textwidth}{!}
{\includegraphics{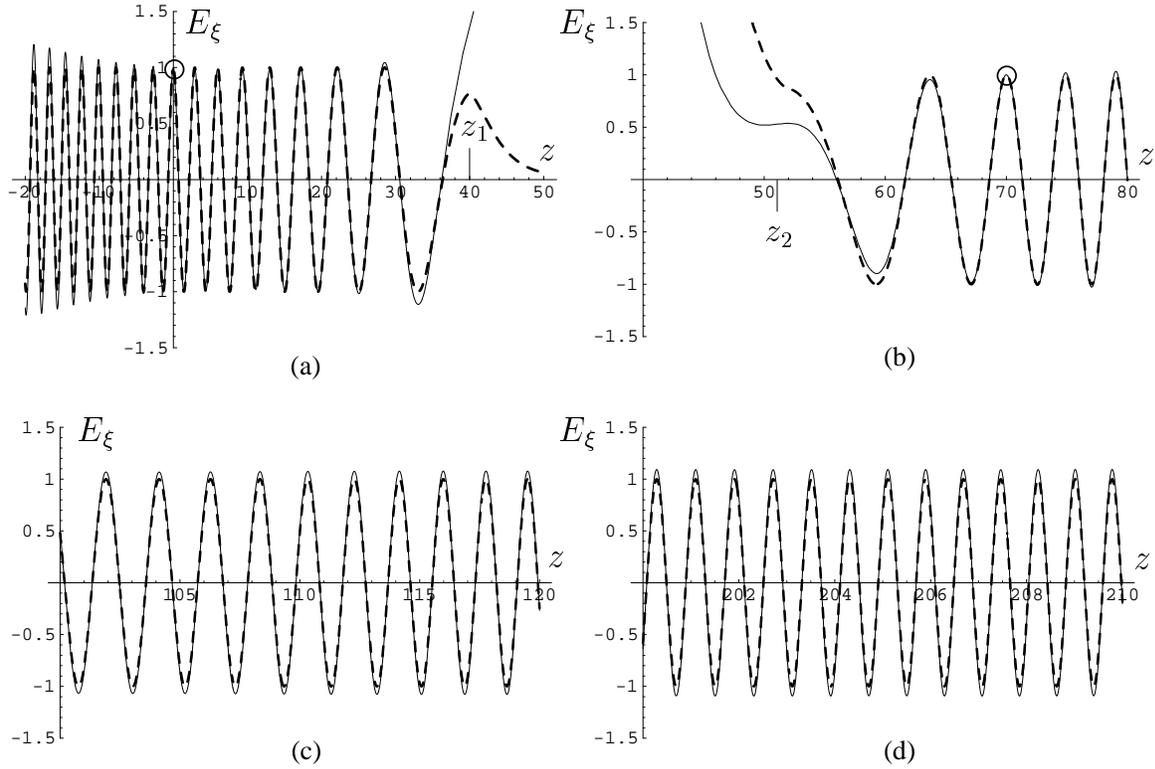}} \caption{Comparison of the full
numerical solution (continuous thin line) and the approximate
solution (overlaid dashed line): the approximate solution is based
on an numerical integration of the local de Vries expression
(\ref{approx}) and eq.~(\ref{wkb}). In fig. (a), the numerical
solution is forced to have the same initial conditions at the
point $z_0=0$, whereas in figs. (b)--(d) $z_0=70$ (circled). Plots
(c) and (d) demonstrate the stability of the approximate solution
for values of positions $z$ distant from $z_0$. Note that the
$z$-range is different in each of the diagrams (b)--(d) indicating
faster oscillations of the field $(E_{\xi},E_{\eta})$ for large
$z$ in the same way as $q(z)$ increases according to
eq.~(\ref{linear-ramp}).} \label{comp}
\end{figure*}
Figure~\ref{comp} compares the numerical solution with the
approximate solution obtained by using $S_2'$. For this purpose,
we choose the approximate solution $S_2'$ corresponding to the
branch (2) in fig.~\ref{local-wave-vec}, integrate it numerically
and obtain the resulting wave by eq.~(\ref{wkb}).  The numerical
solution is obtained by first evaluating eq.~(\ref{wkb}) at an
anchor point $z_0$ giving the four necessary initial conditions:
the two field components $E_{\xi}$ and $E_{\eta}$, and their
gradients $E_{\xi}'$ and $E_{\eta}'$.  In fig. \ref{comp}~(a), we
chose $z_0=0$, and in figs.~\ref{comp} (b)--(d) $z_0=70$.  We
observe that the agreement between the two solutions breaks down
on approaching $z_1$ and $z_2$.  This has a plausible explanation:
we have seen in fig.~\ref{comparison} that the approximation
$S_2'$ breaks down near the points $z_1$ and $z_2$  as in WKB. We
therefore expect the corresponding solution of the wave equation
that is generated from $S_2'$ to fail dramatically at the same
point.

The discrepancy at $z_1$ and $z_2$ between the approximate and
numerical solution can also be explained from another viewpoint:
between $z_1$ and $z_2$, we ideally expect two fundamental
solutions, the one increasing and the other decreasing in an
exponential-like fashion. As the procedure described in the last
paragraph introduces small errors in the initial conditions at
$z_0$, the numerically computed solution does carry components of
the two fundamental solutions. As $z$ enters the region between
$z_1$ and $z_2$, one solution decays rapidly, but the other
(initially only marginally present) increases quickly, becoming
the dominant contribution in the region $[z_1,z_2]$.  Away from
the gap the agreement is very good -- over large distances above
the gap (indicating great stability) and even backwards to the
point $z=-20$ where locally the pitch is infinite.

\subsection{Transmission coefficient}

One can estimate the dependence of the transmission coefficient on the
principal variables $q_0$, $\gamma$, $k_0$, $\epsilon_1$ and
$\epsilon_2$.

Our system is analogous to the simple quantum mechanical barrier
problem for a free particle (see for example \cite{davydov76};
here we have a vector field $\vec{E}$ with two components rather
than one): outside a region $[0,l]$, propagation is described by
the Schr\"odinger equation of a {\it free} particle with an
oscillating wave function and allowing for an incoming and an
outgoing wave.

However, in the region $[0,l]$, the particle moves across a
potential $V$. If the potential $V$ is constant and greater than
the particle's energy $E$, the solutions to the wave equation are
either exponentially decaying or increasing.  The reflection or
transmission coefficients can easily be calculated exactly.
 For a general potential barrier $V(z)$, the transmission coefficient  can be estimated
\cite{davydov76}:
$$
    T\approx
       \exp\left(-2 \frac{1}{\hbar}\int_{0}^{l}dz
        \sqrt{2m(V(z)-E)}\right).
$$

We estimate the transmission coefficient $T$ as the square of the
ratio of the fields $E_{\xi}$ or $E_{\eta}$ at $z_1$ and $z_2$:
\begin{eqnarray}
    T\approx
    \left|\frac{E_{\eta}(z_2)}{E_{\eta}(z_1)}\right|^2 \approx
    \left|\frac{E_{\xi}(z_2)}{E_{\xi}(z_2)}\right|^2
\approx
    \left|\frac{e^{iS(z_2)}}{e^{iS(z_1)}}\right|^2 \nonumber
\end{eqnarray} where the latter form arises from our
assumption~(\ref{wkb}).  The barrier form of $T$ now yields
\begin{eqnarray}
T &=& \Big|\exp\big(
        2i\big[S(z_2)-S(z_1)\big]\big)\Big|\nonumber \\
&&\approx
    \exp\left(2i\int_{z_1}^{z_2}
        dz \ S_2'(z)\right),\label{trans}
\end{eqnarray}
where we have implicitly assumed that the integral of $S_2'(z)$
over $z$ between $z_1$ and $z_2$ is purely imaginary, yielding a
real number in the final exponential of eq.~(\ref{trans}). In the
forbidden zone $[z_1,z_2]$, there are two solutions for $S'_2$
(see. eq.~(\ref{approx})); both are purely imaginary, but of
opposite signs. Therefore, we have to choose the correct one to be
inserted into eq.~(\ref{trans}), corresponding to the decaying
wave.

\begin{figure}[t]
\centering \resizebox{0.4\textwidth}{!}
{\includegraphics{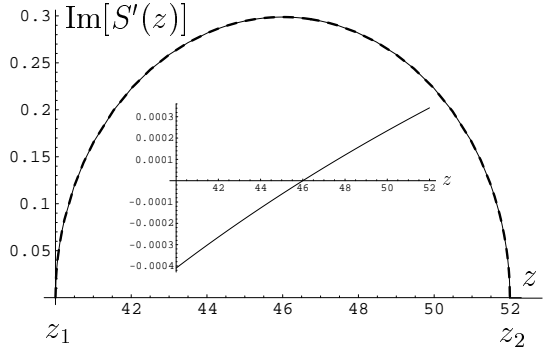}} \caption{Quality of the approximation
(\ref{ellipse-approx}) to model the function $S_2'$ in the evanescent
region between $z_1$ and $z_2$: the thin curve corresponds to the
imaginary part of $S_2'$, whilst the overlaid dashed curve is the
approximation according to eq.~(\ref{ellipse-approx}). The inset shows
the relative error.}
\label{im-comp}
\end{figure}

The function $S_2'(z)$ in (\ref{approx}) has divergent slopes at
the boundary points $z_2$ and $z_1$ and assumes a maximum in
between with
$$
    S'_{2\,{\rm max}} =\frac{\epsilon_2-\epsilon_1}
    {\sqrt{8}\sqrt{\epsilon_1+\epsilon_2}}k_0.
$$
Therefore one might approximate $S_2'(z)$ by an ellipse between the
points $z_1$ and $z_2$ with a maximum value of  $S'_{2\,{\rm max}}$:
\begin{eqnarray}
    S'_2(z) & \approx & S'_{2\,{\rm app}}(z) \label{ellipse-approx}\\
    & = & \frac{1}{\sqrt{2}}
        \frac{ik_0}{\sqrt{\epsilon_1+\epsilon_2}}
        \sqrt{\big(q-\sqrt{\epsilon_1 } k_0\big)
                \big(\sqrt{\epsilon_2} k_0-q\big)}\; ,\nonumber
\end{eqnarray}
where $q=q_0+q_0^2\gamma z$. Figure \ref{im-comp} shows the
$S'_2(z)$ and overlaid the approximation $S'_{2\,{\rm app}}$ of
eq.~(\ref{ellipse-approx}). The inset shows the relative error
$(S'_{2}-S'_{2\,{\rm app}})/S'_{2}$.

Finally, this expression can be integrated analytically to give an
estimate for the integral in eq.~(\ref{trans}), which corresponds
to the area under the curve in fig.~\ref{im-comp}. We obtain
\begin{eqnarray}
    T&\approx&
    \exp\left(
    -\frac{\pi}{\sqrt{2}\,4}\frac{k_0^2}{q_0^2 \gamma}
    \frac{\left(\sqrt{\epsilon_1}-\sqrt{\epsilon_2}\right)^2}
        {\sqrt{\epsilon_1+\epsilon_2}}
    \right)\nonumber\\
    &=&
    \exp\left(
    -\frac{\pi}{\sqrt{2}\,4}
    \frac{q_0^2\gamma \left(z_1-z_2\right)^2}
        {\sqrt{\epsilon_1+\epsilon_2}}
    \right)\label{trans-coeff}
\end{eqnarray}
using eq.~(\ref {space-gap}) for the scaling of $z_1 - z_2$.

To arrive at this estimate (\ref{trans-coeff}) of $T$, we used the
approximation $S'_{2}(z)$, which breaks down near the edges of the
stop gap at $z_1$ and $z_2$, as fig. \ref{comparison} shows. However,
as one can see from fig. \ref{im-comp}, the dominant behaviour of $T$
is determined by the large imaginary values of $S'_{2}$. These occur
deep inside the stop gap where the approximations are valid.

Our expectation at eq.~(\ref {space-gap}) has been confirmed: the
wider the spatial extent of the stop gap the smaller the penetration
and hence the lower the transmission.  Small $\gamma$, that is slow
spatial variation of the band structure, gives lower transmission
(however providing that the sample is thicker than the spatial extent
of the gap!).  One requires $\gamma\ll (n_2-n_1)^2(k_0/q_0)^2$ for
effective reflection at the effective frequency $k_0$.

\section{Conclusions}
The numerical solution of the field amplitudes for light in a
spatially varying cholesteric (a photonic solid) has been
presented. Fields decay in the region where, for the relevant
frequency of light, a uniform cholesteric would have its stop gap.
Transmission is low for sufficiently slow variation of the pitch.
Thus the simple idea that each component of a broad spectrum of
light is reflected where its Bragg (strictly its de Vries)
condition is satisfied, is partially borne out.

Approximate field amplitudes are derived by taking this local de
Vries condition literally.  They are reasonable approximations far
from the gap, but less good at the gap, for reasons we discuss.

Slower spatial variation, $\gamma$ of pitch gives much lower
transmission, in fact exponentially lower, varying as
$e^{-A/\gamma}$. This leads to better utility as broad band
reflectors in devices.

\begin{acknowledgement}
We would like to thank Peter Haynes for inspiring discussions,
especially regarding the interpretation of numerical results and
WKB. S.K.
 acknowledges the financial support of EPSRC.
\end{acknowledgement}

\end{document}